\documentclass[aps,showpacs,twocolumn,showkeys,amsmath,amssymb,floatfix,nofootinbib,preprintnumbers]
{revtex4}
\usepackage{bm}

\begin{document}

\title{How Often Do Diquarks Form?  A Very Simple Model}

\author{Richard F. Lebed}
\email{richard.lebed@asu.edu}
\affiliation{Department of Physics, Arizona State University, Tempe,
Arizona 85287-1504, USA}

\date{June, 2016}

\begin{abstract}
  Starting from a textbook result, the nearest-neighbor distribution
  of particles in an ideal gas, we develop estimates for the
  probability with which quarks $q$ in a mixed $q$, $\bar q$ gas are
  more strongly attracted to the nearest $q$, potentially forming a
  diquark, than to the nearest $\bar q$.  Generic probabilities lie in
  the range of tens of percent, with values in the several percent
  range even under extreme assumptions favoring $q\bar q$ over $qq$
  attraction.
\end{abstract}


\maketitle


\section{Introduction} \label{sec:Intro}

The observation of multiple heavy-quark exotics in recent years,
starting with the Belle discovery of the presumptive $qq \bar q \bar
q$ state $X(3872)$ in 2003~\cite{Choi:2003ue}, has provided entirely
new opportunities for developing a deeper understanding of the QCD
dynamics responsible for binding quarks into color-singlet hadrons.
While it is mathematically true that all SU(3)$_c$ color singlets
assembled from quarks can be decomposed as products of the $qqq$ and
$q\bar q$ combinations familiar from conventional baryons and mesons,
respectively,\footnote{If valence gluons $g$ are included, then
hybrids $q\bar q g$ and glueballs $gg$, $ggg$, {\it etc.} complete the
list of color-singlet substructures.} group theory alone does not
dictate the nature of structures dynamically generated within hadrons.

Several theoretical pictures have been advocated to describe the
multiquark exotics.  The mathematical feature of SU(3)$_c$ just
described encourages one to consider {\em molecules\/} of
color-singlet hadrons, and heavy-quark models of this sort have been
contemplated for almost the entire history of
QCD~\cite{Voloshin:1976ap,DeRujula:1976zlg}.  Alternately, in {\em
hadroquarkonium}~\cite{Voloshin:2007dx}, the heavy $q\bar q$ pair lies
at the center of a cloud generated by the lighter quarks.  In {\em
diquark\/} models, popularized for light-quark systems in
Refs.~\cite{Jaffe:1976ih,Jaffe:1978bu} and for the new heavy
quarkoniumlike exotics like $X(3872)$ in Ref.~\cite{Maiani:2004vq},
diquark $qq$ and $\bar q \bar q$ pairs form via the color attractive
channel ${\bf 3} \times {\bf 3} \to \bar{\bf 3}$ and its conjugate, as
discussed below.  The {\em dynamical diquark\/}
picture~\cite{Brodsky:2014xia,Lebed:2015tna} further purports that the
diquarks do not act as components of stable molecules, but rapidly
separate until confinenment forces the system to hadronize.  And in
{\em kinematic-effect\/} models, first suggested for the new exotic
states in Ref.~\cite{Bugg:2004rk}, the opening of hadronic thresholds
can generate structures resembling resonances nearby in mass, either
by themselves or by coupling to other channels.

In this work we are interested in diquarks, and particularly the
relative rate at which $qq$ (or $\bar q \bar q$) pairs form compared
to the rate for $q\bar q$ pairs.  There exists, after all, no
universally accepted experimental evidence for the existence of
diquarks, so that one might suspect their formation to be a rather
rare occurrence.  Nevertheless, fundamental QCD considerations suggest
otherwise.  The color dependence of the short-distance coupling of
elementary particles in SU(3)$_c$ representations $R_1$ and $R_2$ to
the product representation $R$ is proportional to the combination
\begin{equation} \label{eq:CasimirDef}
{\cal C}(R,R_1,R_2) \equiv C_2 (R) - C_2 (R_1) - C_2 (R_2) \, ,
\end{equation}
where $C_2$ is the representation's quadratic Casimir.  For $qq$ or $q
\bar q$ systems, one finds the relative size of the couplings from
Eq.~(\ref{eq:CasimirDef}) to be
\begin{equation} \label{eq:CasimirNums}
{\cal C}(R,R_1,R_2) =  \frac 1 3 (-8,-4,+2,+1) \ {\rm for} \ R =
({\bf 1},\bar {\bf 3},{\bf 6},{\bf 8}) \, ,
\end{equation}
respectively.  As one might expect, the strongest coupling is that of
the color-singlet $q\bar q$ combination, which provides a direct route
for the formation of mesons.  However, the aforementioned diquark
coupling is a full one-half as strong at short distance, while the two
repulsive channels are both smaller than either of the attractive
ones.

Still, one might expect that in a $qq\bar q\bar q$ system, even if
diquarks initially form, the greater attraction in the $q\bar q$
singlet channel suggests that the system subsequently rearranges
itself into a two-singlet combination.  This issue is less acute in
the dynamical diquark picture, in which the diquarks achieve
substantial separation before this rearrangement can occur.
Nevertheless, the attraction described by Eq.~(\ref{eq:CasimirNums})
strictly holds only at short distances, where single-gluon exchanges
dominate.  One expects the interaction between a particular $qq$ or
$q\bar q$ pair at separations beyond a few tenths of a fm to be
heavily screened by gluon and sea-quark pair creation.

The question of whether diquarks actually appear as important hadronic
substructures then can be discussed in terms of the exact nature of
the spatial distribution of the nearest neighbors of the quarks in the
production process.  Statistically speaking, with some finite
probability a $q$ will find itself much closer to another $q$ than to
a $\bar q$, which allows the diquark attraction to dominate.  But pure
spatial proximity cannot be the whole story, however, since in weak
and electromagnetic decays the created $q\bar q$ pair automatically
forms a color singlet, and yet does not always by itself form a single
meson (the so-called color-suppressed decay diagrams give one
counterexample); a large initial relative momentum between quarks can
apparently overwhelm the proximity effect in certain circumstances,
and hadronization can be delayed until after the quarks lose a
significant amount of energy.

We therefore attempt to remove such complications and create a toy
model as simple as possible.  We generalize the system of the handful
of quarks and antiquarks created in a typical decay or collider
process---whose numbers are exactly equal in a meson decay, $e^+ e^-$,
or $p\bar p$ collision, or differ only by 6 in a $pp$ collision---as
forming a ($q$,$\bar q$) gas of arbitrarily large extent, and assume
that the particles have become essentially static (or at least achieve
something resembling a low-temperature thermal distribution) prior to
hadronization.  Indeed, we model the system as a two-component ($q$
and $\bar q$) ideal gas and ask a very basic question: With what
probability is a given $q$ preferentially attracted---at least
initially---to another $q$ rather than a $\bar q$?  We show that this
probability could be as large as tens of percent, and indeed is
difficult to reduce to lower than a few percent.  Diquarks should be
common components in hadronic processes.

If one subscribes to the long-studied idea~\cite{Anselmino:1992vg}
that all baryons have a significant diquark component, then the simple
model studied here also provides a first step to addressing the
relative rates of meson, baryon, and tetraquark production.

It is important to point out that other works discussing diquark
production focus on different energy, density, or temperature regimes.
Direct diquark production is built into jet fragmentation event
generators dating back at least as far as the famous Lund
model~\cite{Andersson:1983ia}; however, the diquark attraction
described here is of the lower-energy ``non-prompt'' variety.  Diquark
condensation in dense QCD is a well-studied phenomenon ({\it e.g.},
\cite{Rapp:1997zu}), and has been extended also to finite
temperature~\cite{Liao:2005pa}; obviously, an ideal low-temperature
gas is neither of these.  In addition, in more formal work using an
effective supersymmetric embedding of quantum mechanics into AdS
space~\cite{Brodsky:2016yod,Brodsky:2016rvj}, diquarks are found to be
absolutely natural and indeed essential hadronic components: In
particular, the $qq$ and $q\bar q$ attraction strengths turn out to be
the same, and baryons are naturally quark-diquark bound states.

This paper is organized as follows.  In Sec.~\ref{sec:Ideal}, the
problem of nearest neighbors in an ideal gas is treated in both the
original textbook case and the two-component case.
Section~\ref{sec:Model} applies these ideas to the case of diquark
attraction, and we outline the many reasons why this treatment falls
far short of real QCD, as well as the ways in which the model attempts
to address at least some of them.  Explicit model calculations appear
in Sec.~\ref{sec:Results}, where we find that diquark attraction
should be a rather common occurrence in hadronic physics.
Section~\ref{sec:Concl} summarizes and concludes.

\section{Nearest-Neighbor Distances in Ideal Gases}
\label{sec:Ideal}

\subsection{The Classic Problem}

We begin with a standard textbook problem, that of the distribution of
nearest-neighbor particles randomly (Poisson) distributed to form an
ideal gas.  This problem was first addressed by P.~Hertz in
1909~\cite{Hertz:1909}.  We present here the elegant derivation by
Chandrasekhar~\cite{Chandrasekhar:1943ws}, as it is useful both for
establishing notation and for being amenable to straightforward
generalizations.

In the original problem, the gas particles are classical, pointlike,
and noninteracting except through possible elastic collisions, and
(implicitly) obey Maxwell-Boltzmann statistics.  The system
effectively is infinite in extent and isotropic, so that any point may
be treated as typical.

We are interested in the radial probability density $w(r)$ of the
nearest particle to the (arbitrary) origin to lie at a distance $r$,
which satisfies the normalization condition
\begin{equation} \label{eq:Norm}
\int_0^\infty dr \, w(r) = 1 \, ,
\end{equation}
noting that, in light of the isotropy assumption, the angular
integrals and volume-element $r^2$ have already been absorbed into the
definition of $w(r)$.  We are also interested in the mean
nearest-neighbor distance,
\begin{equation} \label{eq:Meanr}
\left< r \right> \equiv \int_0^\infty dr \, r w(r) \, .
\end{equation}
Let $n$ be the volume number density of particles.  Then $w(r)$
satisfies
\begin{equation} \label{eq:IntEq}
w(r) dr = \left[ \int_r^\infty \! dr^\prime w(r^\prime) \right] n
\cdot 4\pi r^2 dr \, .
\end{equation}
In words, Eq.~(\ref{eq:IntEq}) says: The probability for the nearest
neighbor to the origin to lie in a spherical shell of radius $dr$ at
$r$ [which is $w(r)dr$] equals the product of the probability that no
particle lies between the origin and radius $r$---{\it i.e.}, that the
nearest neighbor lies between $r$ and $\infty$ (which is the
integral)---times the number of particles in the spherical shell
(which is $n \cdot 4\pi r^2 dr$).  Equation~(\ref{eq:IntEq}) may be
recast as a differential equation,
\begin{equation} \label{eq:DiffEq}
\frac{d}{dr} \left[ \frac{w(r)}{4\pi r^2 n} \right] = -4\pi r^2 n
\left[ \frac{w(r)}{4\pi r^2 n} \right] \, .
\end{equation}

The original problem assumes $n$ constant (appropriate to a uniform
ideal gas of infinite extent), but for later use, let us generalize to
the case of a radially dependent density $n(r)$ (which implies a
unique origin $r=0$ for the system).  The solution to
Eq.~(\ref{eq:DiffEq}) is
\begin{equation} \label{eq:ExactSoln}
w(r) = C \cdot 4\pi r^2 n(r) \exp \left[ -4\pi \int_0^r dr^\prime \,
n(r^\prime) r^{\prime \, 2} \right] \, .
\end{equation}
The integration constant $C$ is fixed by imposing the normalization
condition Eq.~(\ref{eq:Norm}).  One finds
\begin{equation} \label{eq:IntConst}
C = \left\{ 1 - \exp \left[ -4 \pi \int_0^\infty dr \, n(r) r^2
\right] \right\}^{-1} .
\end{equation}
In the case of constant $n$, or at least $n(r)$ that remains
sufficiently large as $r \to \infty$, the exponential approaches zero,
and $C \to 1$.

In the original case of constant $n$, the integrals may be performed
analytically, giving
\begin{eqnarray}
w(r) & = & 4 \pi r^2 n \exp \! \left( \! -\frac{4\pi r^3}{3} n \!
\right) = \frac{3}{a} \! \left( \frac{r}{a} \right)^2 \! \exp \!
\left[ -\frac{r^3}{a^3} \right] , \; \; \; \label{eq:ConstDens} \\
\left< r \right> & = & \Gamma \left( \frac{4}{3} \right) \left(
\frac{4\pi n}{3} \right)^{-\frac 1 3} = \Gamma \left( \frac{4}{3}
\right) a \simeq 0.8928 a \, ,
\end{eqnarray}
where the density $n$ defines the natural length scale, $a \equiv
(4\pi n/3)^{-1/3}$.  Indeed, one may also show that  $\left< r^3
\right> = a^3$.

\subsection{Two-Component Ideal Gas}

We now extend the previous derivation to solve another simple
classical problem, which to our knowledge has not previously been
addressed: Consider an ideal gas consisting of two species, 1 and 2,
with corresponding number densities per unit volume (natural length
scales) $n_i$ ($a_i$), $i = 1, 2$.  Starting from some fiducial point
(the origin), what is the probability that the nearest particle of
type 2 does not appear until a distance $k$ times further than that at
which the nearest particle of type 1 appears?  We have in mind of
course that the origin contains a quark, and want to know the
statistical likelihood that the nearest neighbor happens to be a quark
rather than an antiquark, by a chosen distance ratio $k$.  Note that
the ratio $k$ can lie anywhere in $(0,\infty)$, and that the
specification of the species at the origin ($q$ or $\bar q$) need only
be made at the end of the calculation.

The probability for the first particle of type $i$ lying at a distance
$r_i$ from the origin is $w_i(r_i) dr_i$, where $w_i$ is simply the
function in Eq.~(\ref{eq:ExactSoln}) or (\ref{eq:ConstDens}) defined
with $n \to n_i$.  Since the gas is ideal, the probabilities are
independent, and the combined probability that the first particle of
type 1 lies at distance $r_1$ and the first particle of type 2 lies at
distance $r_2$ from the origin is
\begin{equation}
p(r_1,r_2) = w_1 (r_1) dr_1 \, w_2 (r_2) dr_2 \, .
\end{equation}
Next, the probability $P(r_1,k)$ that the first particle of type 1
lies at a distance $r_1$ from the origin and that the first particle
of type 2 lies at least a distance $kr_1$ from the origin is then
\begin{equation}
P(r_1,k) = \int_{r_2 = kr_1}^\infty p(r_1, r_2) = w_1(r_1) dr_1
\int_{kr_1}^{\infty} w_2(r_2) dr_2 \, ,
\end{equation}
and the probability $P_{1,2}(k)$ that the first particle of type 2
lies at a distance $k$ times that of the first particle of type 1,
regardless of the specific value of $r_1$, is
\begin{equation} \label{eq:Pexact}
P_{1,2}(k) = \int_{r_1 = 0}^\infty P(r_1,k) = \int_0^\infty \! \! w_1
(r_1) dr_1 \int_{kr_1}^\infty w_2 (r_2) dr_2 \, .
\end{equation}
This expression, with inputs suitably chosen, is used in this paper to
calculate likelihoods relevant to the formation of diquarks versus
color-singlet $q\bar q$ pairs.

In the case of constant densities $n_i$, the integrals in
Eq.~(\ref{eq:Pexact}) can be performed in closed form.  Since
\begin{equation}
\int_{R}^\infty dr_i \, w_i(r_i) = e^{-(R/a_i)^3} \, ,
\end{equation}
one finds
\begin{equation} \label{eq:Prob2}
P_{1,2}(k) = \frac{a_2^3}{k^3 a_1^3 + a_2^3} = \frac{n_1}{k^3 n_2 +
n_1} \, .
\end{equation}
Several limits of this simple expression are easy to understand.  It
must vanish as $k \to \infty$, which is the unlikely case that all
type-2 particles are arbitrarily far from the origin; indeed, the
scaling with $1/k^3$ is expected from the volume effect of scaling the
distance ratio as $k$.  The limit $k \to 0$ simply means the case
where the nearest type-2 particle can be anywhere, for which
$P_{1,2}(k)$ must approach unity.  The case $k = 1$ means the
probability of finding the first type-2 particle to lie at least as
far from the origin as the first type-1 particle; if $n_1 = n_2$, one
expects the probabilities for either of the species to provide the
nearest particle to the origin to be equal: $P_{1,2}(1) = \frac 1 2$,
exactly as one finds from Eq.~(\ref{eq:Prob2}).

\section{Idealized Diquark Attraction}
\label{sec:Model}

\subsection{Warmup: Electric Charges}

Using the analysis of the two-component ideal gas in the previous
section, let us begin by considering an analogous problem with static
electric charges.  Of course, even the particles of an ideal gas at
finite temperature have a nontrivial velocity distribution, meaning
that the static assumption is already suspect, and Earnshaw's theorem
moreover forbids such a static system from being in a stable
equilibrium.  Nevertheless, one can take the system at the initial
time $t=0$ to start from rest and assume that the dominant interaction
between charges is the central Coulomb force.  The model two-component
gas consists of type-1 particles of charge +$\frac 1 2$ and type-2
particles of charge +1, designed to emulate the factor-2 difference
between the short-distance strength of the $qq$ and $q\bar q$
channels.  A negative test charge, attracted to both charge species,
is placed at the origin.  How frequently does its initial attraction
to a type-1 particle equal or exceed that to a type-2 particle, due
only to the initial distribution of the particles?

Several limitations of addressing the problem in this way have already
been noted, but let us remark in addition that initial attraction to a
neighbor is not the same as the formation of a compact state with this
neighbor.  The collective effect of several other neighbors can
overwhelm it, the $t \! > \!0$ migration of the test charge towards
its most attractive neighbor disturbs the initial configuration (as
does the movement of the other charges), allowing other charges to
disrupt the initial attraction, and of course the proper treatment of
charges in motion requires one to include the effects of magnetic
fields.

Nevertheless, the idealized problem is well defined.  Since the
Coulomb attraction obeys an inverse-square law, a type-1 particle
provides the most attractive initial interaction for $k=\sqrt{2}$
({\it i.e.}, the nearest type-2 particle lies at least $\sqrt{2}$
times farther from the origin than the nearest type-1 particle).
Assuming that the densities of the two types of particles are equal
and constant, Eq.~(\ref{eq:Prob2}) gives that the test charge is
initially more attracted to the smaller type-1 charge with probability
$P_{1,2}(\sqrt{2}) = 1/(2\sqrt{2}+1) \simeq 26\%$, a sizeable
fraction.

\subsection{Unscreened Quarks}

An ideal, initially stable distribution of $q$ and $\bar q$ is much
more complicated than the example just discussed for several reasons.
First, only the bare (short-distance) quark interaction obeys the
simple scaling from Eq.~(\ref{eq:CasimirNums}) in which the attractive
$qq$ channel has half the strength of the attractive $q\bar q$
channel.  At larger distances, the interactions represented by the
exchange of colored gluons (and additional $q\bar q$ pair creation)
serve to screen the bare interaction; we attempt a simpleminded
modeling of this effect below.  It is worth mentioning that this
``Casimir scaling'' given by Eq.~(\ref{eq:CasimirDef}) is violated
only at three-loop perturbative order~\cite{Anzai:2010td} and its
existence to as much as $r \sim 1$~fm is well supported in lattice
simulations~\cite{Bali:2000un}.

Second, as noted in the Introduction, the short-distance interactions
also feature repulsive $qq$({\bf 6}) and $q\bar q$({\bf 8})
combinations.  One could certainly extend the derivation of the
previous section to a 3- or 4-component gas, but one sees from
Eq.~(\ref{eq:CasimirNums}) that the short-distance $qq$-{\bf 6}
repulsion is only $\frac 1 2$ as large as the $\bar{\bf 3}$
attraction, and the $q\bar q$-{\bf 8} repulsion is only $\frac 1 8$ as
large as the {\bf 1} attraction.  Moreover, we are interested in the
attractive forces that ultimately lead to quarks combining into
hadrons, and therefore ignore the effect of the repulsive channels
(which, presumably, lead to the separation of quark clusters into
hadrons).  However, even under the assumption of neglecting repulsive
forces, these channels have an important effect: A generic $qq$ has 9
possible color combinations, of which only the 3 forming the $\bar{\bf
3}$ are attractive, while for the 9 possible color combinations of a
generic $q\bar q$ pair, only the singlet {\bf 1} combination is
attractive.  If one assumes equal densities of $q$ and $\bar q$, then
the {\em effective\/} density of attractive $q$'s is 3 times the
effective density of attractive $\bar q$'s, {\it i.e.}, $n_1 = 3n_2$.
Again using $k = \sqrt{2}$, one finds from Eq.~(\ref{eq:Prob2}) the
remarkable result
\begin{equation} \label{eq:Unscreened}
P_{1,2} (n_1 = 3n_2; k = \sqrt{2}) = 3(3-\sqrt{2}) \simeq 51.5\% \, ,
\end{equation}
suggesting a very substantial probability for diquark attraction in
the case of unscreened quark color charges.

The assumption that the initial interaction is dominated by the static
color-Coulomb force relies on the nonrelativistic expansion of the
quark bilinears $\bar q \gamma^\mu q \to \delta^\mu_0 q^\dagger q$ in
the QCD Lagrangian; nonzero spatial momenta couple to $\gamma^i$ and
thus induce significant spin dependence in the interaction, especially
for relativistic quarks.  Relativity is also important for corrections
to the assumption of a quark gas of infinite extent, since a fully
correct treatment must include the retardation of the propagating
interactions.  On the other hand, since the exchange symmetry of
$\bar{\bf 3}$ ({\bf 6}) is antisymmetric (symmetric), the effect of
the Pauli exclusion principle and neglecting the {\bf 6} serves to
exclude $qq$ pairs in overall flavor-spin-space symmetric
combinations.  For distinct light ($u$, $d$) quarks in relative
$s$-waves, isosinglet/spin-singlet and isotriplet/spin-triplet pairs
survive this sorting, while for identical light quarks, the first
combination is also excluded.

\subsection{Color Screening}

Despite the significant number of exceptions and corrections to the
ideal gas model thus far identified, the effects discussed above are
still essentially classical.  Real QCD is of course a
quantum-mechanical theory, meaning that the concept of pointlike
particles with potential interactions depending predominantly upon
their separation, and hence the very concept of nearest neighbors,
should be considered suspect.  Quark wave functions have a finite
extent; therefore, one's first thought may be to reanalyze the
nearest-neighbor derivation to include finite particle radii.  A
relevant calculation has been undertaken in Ref.~\cite{Torquato:1990}
that models the particles as hard spheres, and presumably could be
generalized to the case in which the spheres are partially penetrable
``clouds''.  In the context of the model here, such an effect could be
incorporated by altering the functional dependence of the density
functions $n_{1,2}(r)$.  But even these modifications do not fully
respect a fundamental feature of quantum mechanics: Interactions can
collapse wave functions, which is natural in light of the fact that
the fundamental QCD interaction $\bar q(x) \gamma^\mu g_s A_\mu (x)
q(x)$ is local.  We therefore argue that the concept of nearest
neighbors retains its significance when interpreted in the usual
statistical sense of quantum mechanics.

The most important quantum-mechanical effect in modifying the
ideal-gas picture, however, is a quantum field-theoretical effect: the
crucial importance of color screening in strong interactions.  The
large size of the strong coupling $\alpha_s = g_s^2/4\pi$ at low
energies means large numbers of sea-quark $q\bar q$ pairs created
between the original quarks of the ideal gas, which serve to screen
the initial color interactions.  Of equal significance, the
non-Abelian nature of QCD means that the gluons themselves carry color
charge, and hence self-couple and contribute to the color screening.
Were the exact solutions to the renormalization group equations known
at low scales, a rigorous treatment of the screening could be
undertaken.  In practical terms, the dominant feature of low-energy
QCD is color confinement, whose effect we incorporate as an effective
screening of color charges beyond a radius given by the largest
typical hadron sizes, $R = O$(1~fm).

Before examining the results of explicit model calculations, we
summarize the multiple roles played by the density functions
$n_{1,2}(r)$.  We have used their overall scales to represent not only
the relative densities of $q$ and $\bar q$, respectively, but also the
relative number of channels attractive to a test quark at the origin.
Their functional dependences could in principle be used to model the
finite extent of $q$, $\bar q$ wave functions, but here we use it to
model the strong color screening in a $q$, $\bar q$ gas by introducing
a characteristic screening radius $R$.  The only other independent
parameter in the model is the ratio $k$, indicating the relative
distance at which a $\bar q$ and $q$ are attracted to the test quark
at the origin with equal strength.

\vspace{-1ex}
\section{Explicit Models} \label{sec:Results}

Starting with the expressions Eqs.~(\ref{eq:ExactSoln}),
(\ref{eq:IntConst}), and (\ref{eq:Pexact}), we model the effective
screened densities using several plausible functional forms, and
investigate the results for the probabilities $P_{1,2}(k)$ as an
indication of the likelihood of diquark attraction.

The three functional profiles are all chosen to have $n_1 = 3n_2$ as
discussed above, although altering the specific ratio of 3 does not
alter the ultimate significance of the substantial values obtained for
$P_{1,2}(k)$, as discussed below.  The profiles are a hard-wall
screen,
\begin{equation} \label{eq:Profile1}
n_1^{(1)} = n_0 \, \Theta( R - r) \, ,
\end{equation}
where $\Theta$ is the Heaviside step function; a Saxon-Woods form with
a skin depth $d$,
\begin{equation} \label{eq:Profile2}
n_1^{(2)} = n_0 \cdot \frac{1 + {\rm exp} \left( -\frac{R}{d} \right)}
{1 + {\rm exp} \left( \frac{r-R}{d} \right)} \, ;
\end{equation}
and a linear decrease out to the screening wall at $R$,
\begin{equation} \label{eq:Profile3}
n_1^{(3)} = n_0 \left( 1- \frac{r}{R} \right) \Theta( R - r) \, ;
\end{equation}
all of which have the same central unscreened density $n_0$.  For
definiteness, we choose $n_0 = (2/R)^3$ (indicating an expectation of
encountering two $q$'s or $\bar q$'s before reaching $R$) and $d =
R/2$.  We compare results for the three profiles and explore their $k$
dependence in Table~\ref{tab:Varyk}.  Note first that the result for
the hard-wall screen $n^{(1)}$ with $k = \sqrt{2}$ and $n_1 = 3n_2$
almost equals the result of Eq.~(\ref{eq:Unscreened}), meaning that
the color screening has little effect when $k = \sqrt{2}$ and $n_0 =
(2/R)^3$.  In fact, Eq.~(\ref{eq:Unscreened}) with larger values of
$k$ continues to match the results in the first line of
Table~\ref{tab:Varyk} quite well, within 10\%.  The Saxon-Woods form
$n^{(2)}$ with a substantial skin depth $d=R/2$ actually gives
somewhat larger values of $P_{1,2}(k)$ than the hard-wall form
$n^{(1)}$, due to the sampling of points for $n^{(2)}$ with $r > R$.
Of course, in the limit $d \to 0$, its profile Eq.~(\ref{eq:Profile2})
reduces to that of Eq.~(\ref{eq:Profile1}), which can also be checked
numerically.  Even a profile like $n^{(3)}$ in
Eq.~(\ref{eq:Profile3}), for which the effective density decreases to
zero at $r = R$, decreases the values of $P_{1,2}$ somewhat but leaves
their order of magnitude intact.  One can check that enhancing the
rate of decrease even to a profile $n(r)$ that falls exponentially
fast (not exhibited here) does not fundamentally change this
conclusion.
\begin{table}
\caption{Values of $P_{1,2}(k)$ in percent for the three screened
density profiles of Eqs.~(\ref{eq:Profile1})--(\ref{eq:Profile3}), for
three values of $k$.}
\label{tab:Varyk}
\begin{tabular}{c|rrr}
\hline\hline
& \ \ $k = \sqrt{2}$ & \ \ $k = 2\sqrt{2}$ & \ \ $k = 3\sqrt{2}$ \\
\hline
$n_1^{(1)}$ & 50.29 & 10.85 &  3.46 \\
$n_1^{(2)}$ & 53.24 & 13.99 &  4.79 \\
$n_1^{(3)}$ & 40.40 &  9.04 &  3.04 \\
\hline\hline
\end{tabular}
\end{table}

Besides the value of $k$ and the shape of the profile function $n(r)$,
the only remaining degrees of freedom in this model are the precise
value of the central density $n_0$ and the relative ratio $n_1/n_2$ of
$q$ to $\bar q$ channels attractive to the central quark, which was
argued in the previous section to be 3.  The value of $n_0$ affects
the results of the calculation due to its appearance in the
exponentials of Eqs.~(\ref{eq:ExactSoln})--(\ref{eq:IntConst}); the
main effect of changing its value is to change its precise
relationship to the length scales in the problem, particularly to the
screening radius $R$.  For definiteness, consider the hard-wall
profile $n^{(1)}$ and $k = \sqrt{2}$.  The results for several values
of $n_0$ are presented in Table~\ref{tab:Varyn0}.  We see that
increasing $n_0$ results in a rapid approach to the unscreened case of
Eq.~(\ref{eq:Unscreened}); and even a decrease of $n_0$ to $1/(10R)^3$
only results in a decrease of $P_{1,2}(k)$ by a factor of 3, indeed
with values approaching the asymptotic value $1/2k^3$ (when $k > 1$),
as can be shown analytically from Eqs.~(\ref{eq:ExactSoln}),
(\ref{eq:IntConst}), and (\ref{eq:Pexact}).
\begin{table}
\caption{Values of $P_{1,2}(k = \sqrt{2})$ in percent for the screened
density profile $n_1^{(0)}$ of Eq.~(\ref{eq:Profile1}), for several
values of $n_0$.}
\label{tab:Varyn0}
\begin{tabular}{c|cccccc}
\hline\hline
$n_0$ & $1/(10R)^3$ & $1/(5R)^3$ & $1/(2R)^3$ & $1/R^3$ & $(2/R)^3$ &
$(3/R)^3$ \\
\hline
$P_{1,2}(\sqrt{2})$ & 17.69 & 17.74 & 18.69 & 26.01 & 50.29 & 51.47 \\
\hline\hline
\end{tabular}
\end{table}

Lastly, with respect to the ratio $n_1/n_2$, one can check that the
hard-wall density profile $n^{(1)}$ of Eq.~(\ref{eq:Profile1}) gives
results almost exactly matching the idealized unscreened formula
Eq.~(\ref{eq:Prob2}).  For example, if $n_1/n_2$ is not 3 as suggested
by simple color considerations, but somehow the effective density of
attractive $\bar q$'s is 10 times this value, then $P_{1,2}$ reduces
from 50.29\% to 9.59\%, still a rather significant value.  In every
case, the probability $P_{1,2}$ of a test quark being initially
attracted most strongly to another $q$ rather than a $\bar q$ is at
least several percent.

\section{Conclusions} \label{sec:Concl}

We have seen that the large relative size of the short-distance
attraction between quarks in the color-antitriplet channel compared to
the attraction between a quark and an antiquark in the color-singlet
channel leads inexorably to a given quark being initially attracted to
a quark rather than an antiquark a sizeable fraction of the time.  We
interpret this initial attraction as the seed event in the formation
of a compact diquark $qq$ rather than a color-singlet $q\bar q$ pair.

While the short-distance color attraction must be modified by QCD
renormalization effects and the color-screening effects due to
confinement, one still expects this attraction to extend out to some
finite distance, and at greater distances still if both quarks and
antiquarks are comparably screened.  Under these assumptions, the
probability of preferential $qq$ attraction is in the tens of percent.
Even if one allows the $q\bar q$ attraction to somehow dominate by a
factor of several at these distances (either through the relative
force of attraction, modeled here by the parameter $k$, or through the
relative effective density of $\bar q$ compared to $q$ attracted to
the test quark, modeled here by $n_1/n_2$), the probability of
preferential $qq$ attraction is still at least several percent.  The
key formula describing the most idealized situation, from which
rule-of-thumb estimates may be obtained, is the unscreened result
Eq.~(\ref{eq:Prob2}).  In the context of the class of models described
here, it is very difficult to completely suppress the $qq$ attraction
far below the level of $q\bar q$ attraction.

What evidence, then, does one have of diquark production?  While
tantalizing hints of diquark substructure have appeared in multiple
light-quark systems in the past (such as the $f_0$ and $a_0$ mesons),
the heavy-quark systems offer greater opportunities for disentangling
clear signals because of the well-defined spectroscopy of heavy
quarkonium systems.  Consider one example, the charged $J^P =1^+$
exotic $Z^-(4430)$, confirmed at LHCb at high statistical
significance~\cite{Aaij:2014jqa}, which is a prime candidate for a
diquark-antidiquark system~\cite{Brodsky:2014xia,Maiani:2014aja},
particularly if the diquarks are considered well separated, as in
Ref.~\cite{Brodsky:2014xia}.  Its only measured decay mode thus far is
$Z^-(4430) \to \psi(2S) \pi^-$, and its combined branching fraction in
the decay chain $B^0 \to Z^-(4430) K^+ \to \psi(2S) \pi^- K^+$
is~\cite{Agashe:2014kda} $(6.0^{+3.0}_{-2.4}) \times 10^{-5}$.  In
comparison, the branching fraction of $B^0$ to the much lighter
conventional $1^{++}$ charmonium state $\chi^{\vphantom\dagger}_{c1}$
(3511~MeV) and a $K^0$ is $(3.93 \pm 0.27) \times 10^{-4}$, with
similar values for other two-body charmonium decays.  The branching
fraction to $\chi_{c1}$ is only a factor of 4--12 times larger,
despite a two-body phase space that is 2.4 times larger.  A large
probability for diquark formation under these assumptions seems to be
indicated.

Such positive signals from both data and from this simple model will
hopefully encourage the development of much more realistic QCD
treatments of $qq$ attraction and diquark formation.  It seems
unavoidable that allowing for a substantial degree of diquark
formation will need to be taken into account in future studies of
high-energy and heavy-quark hadronic physics.

\vspace{-3.8ex}
\begin{acknowledgments}
\vspace{-2ex}
  I thank A.D.~Polosa and S.J.~Brodsky for enlightening suggestions.
  This work was supported by the National Science Foundation under
  Grant No.\ PHY-1403891.
\end{acknowledgments}


\end{document}